\documentstyle[preprint,aps,psfig]{revtex} 
\begin{document}
\draft

\title{Interaction effects on persistent current of ballistic cylindrical nanostructures}

\author{St\a'ephane Pleutin}
\address{Physikalisches Institut, Albert-Ludwigs-Universit\"at-Freiburg, Hermann-Herder Stra\ss e 3, D-79104 Freiburg, Germany}

\date{\today}
\maketitle

\begin{abstract}
We consider clean cylindrical nanostructures with an applied longitudinal static magnetic field. Without Coulomb interaction, the field induces, for particular values, points of degeneracy where a change of ground state takes place due to Aharonov-Bohm effect. The Coulomb potential introduces interaction between the electronic configurations. As a consequence, when there is degeneracy, the ground state of the system becomes a many body state - unable to be described by a mean-field theory - and a gap is opened. To study this problem, we propose a variational multireference wave function which goes beyond the Hartree-Fock approximation. Using this ansatz, in addition to the avoided crossing formation, two other effects of the electron-electron interaction are pointed out: (i) the long-range part of the Coulomb potential tends to shift the position in magnetic field of the (avoided) crossing points and, (ii) at the points of (near) degeneracy, the interaction can drive the system from a singlet to a triplet state inducing new real crossing points in the ground state energy curve as function of the field. Such crossings should appear in various experiments as sudden changes in the response of the system (magnetoconductance, magnetopolarisability,...) when the magnetic field is tuned. 

\end{abstract}


\vskip2pc

\section{introduction}

It is known, since the early days of quantum mechanics, that the low energy electronic properties of aromatic molecules are very sensitive to a magnetic field applied perpendicularly to their planes \cite{aromatic}. The field breaks the time reversal symmetry and therefore induces an electronic current running around the circumference of the ring. This is the persistent current, arising due to Aharonov-Bohm effect \cite{ab}. This is an equilibrium phenomenon, periodic in magnetic flux with period $\phi_0=hc/e$, the flux quantum \cite{byers}. But, to be able to measure it requires systems with characteristic lengths in the nanoscopic or mesoscopic scale. Indeed, on the one hand, the motion of the electrons has to stay coherent which is possible for systems smaller than the electronic coherence length. But, on the other hand, to cover a full period of magnetic flux requires rings with sufficiently large diameters. This is definitely not the case for the usual aromatic molecules where field as large as $10^5$ tesla are needed to observe the periodicity. However, nowadays, many man-made systems in the appropriate length scale, are available in various forms: isolated or ensemble of metallic or semiconducting rings \cite{expring} and, more recently, carbon nanotubes which are large organic cylinders \cite{lijima,basel} or, last, rings of carbon nanotubes \cite{avouris,roche}. Therefore, the study of persistent current have regained lot of interest during the last ten years.

Rings, made of usual metal or semiconductor, are studied intensively since the nineties and a few experiments, motivated by early theoretical prediction \cite{buttiker}, have measured a sizable persistent current in some systems \cite{expring}. But, neither the magnitude of the current - one or two orders of magnitude larger than expected - nor the  diamagnetic sign of the response measured experimentally can be explained by existing theories. Most of the theoretical efforts were devoted to the study of the interplay between disorder and Coulomb interaction, but without convincing conclusions up to now \cite{review}. 

Carbon nanotubes were discovered by Iijima in 1991 \cite{lijima}. They are fascinating materials whose electronic properties are determined in a unique way by the topology of their lattice: they are rolled up strip of graphite sheet whose, depending on their diameter and chirality, can be either metallic or semiconducting \cite{saito-book}. In any case, the electronic spectrum of these systems seems to be very sensitive to an applied magnetic field suggesting large orbital magnetic response \cite{ovchinnikov,szopa}. Indeed, recently strong field effects have been seen in measurements of the tunnelling conductance of multiwall carbon nanotubes \cite{basel}. Magnetoconductance measurements of rings of carbon nanotubes have also been done \cite{avouris}.


In order to better understand these experimental results, additional studies of electronic multichannel systems in presence of Coulomb interaction are needed. This is clear for ballistic systems but, even for metallic rings which are in the diffusive regime, a proper understanding of the clean case should help to find a formalism able to treat disorder and interaction on equal footing. In this work, we consider cylindrical systems made of rolled square lattices (multichannel systems), without disorder but with Coulomb potential, long or short range. We are interested in how the Coulomb potential can affect the ground state properties of these systems when a static magnetic field parallel to the cylindrical axis is tuned, and hence, on the behaviours of the persistent current. Similar studies were done in the past but, either for pure 1D systems \cite {kusmartsev} i.e. systems with only one electronic channel (rings), or for multichannel systems (cylinders) but with strong disorder using first order perturbation theory or the Hartree-Fock approximation \cite{bouzerar2}. The conclusions obtained here are not contained in these works. 

For cylinders, in the pure case and without electron-electron interaction, an axial magnetic field induces many level crossings at different values of the magnetic flux \cite{pleutin}. At zero temperature, for a certain density of particles, the ground state energy shows also crossing points for particular values of the field where the ground state is degenerate. With the Coulomb interaction, there exist interaction between these degenerate states resulting in avoided crossings; in other words, at least at the vicinity of the crossing points the ground state becomes a true many-body state unable to be described by any mean field treatment. In this work, we propose a `minimal' variational wave function to deal with this particular problem, going beyond a simple Hartree-Fock calculation. As a result, we find, in addition to the avoided crossing formation, two others effects caused by repulsive interaction. (i) The positions of the (avoided) crossings points are shifted in magnetic field; this effect is due to the non-equal Hartree contributions of the different components of the interacting ground state. (ii) Because of the presence of degeneracy, the Hund's rule may drive the system to a triplet state: it follows sequences of singlet $\rightarrow$ triplet $\rightarrow$ singlet ground states which should be seen in various type of  experiments such as measurements of the magnetoconductance, for instance.

We believe the conclusions of this work rather general providing that the system under consideration is in the ballistic regime i.e. weakly disordered. Indeed, all the effects described here, result from the degeneracy, or quasi-degeneracy,  of the ground state induced by the magnetic field in the case without Coulomb interaction. In principle, this always occurs for any cylindrical systems as carbon nanotubes, ring of carbon nanotubes or rings of usual semiconducting materials.

The paper is organised as follows. In section II, the model without Coulomb interaction is presented and the origin of the crossing points induced by an applied magnetic field is described. In section III, we introduce the model with Coulomb interaction. In a first subsection, exact diagonalisation results are shown and the main interaction effects are discussed. In a second subsection, our variational ansatz is presented and some effects of the Coulomb interaction are shown to be well reproduced by our approximation. Last, in a third subsection, possible spin effects are analysed.

\section{Non interacting electrons. Orbital effects}

The generic systems we consider along this work are rolled square lattices. In this section we start by neglecting the Coulomb interaction to focus on orbital effects only. The electrons are then described by the following nearest-neighbour tight-binding model where an uniform magnetic field, $H$, parallel to the cylindrical axis, is included via the Peierls-London substitution

\begin{equation}
\label{Hcylinder}
\hat{H_0}=t\sum^N_{n=1,\sigma=\pm\frac{1}{2}}(c^{\dagger}_{n+1,m,\sigma}c_{n,m,\sigma} e^{i\frac{2\pi}{N}\phi}+h.c.)+t\sum^{M}_{m=1,\sigma=\pm\frac{1}{2}}(c^{\dagger}_{n,m+1,\sigma}c_{n,m,\sigma}+h.c.).
\end{equation}where $\phi$ is the magnetic flux through the section of the cylinder in units of the flux quantum $\phi_0$ ($\phi_0=hc/e$). The two indices $(n,m)$, two integers, are the coordinates of the lattice sites: $n$ is the coordinate along the circumference, $1\le n \le N$, and $m$ the one along the cylindrical axis, $1 \le m \le M$. The fermionic operator $c^{\dagger}_{n,m,\sigma}$ ($c_{n,m,\sigma}$) is the creation (destruction) operator of an electron at site $(n,m)$ with spin $\sigma$. The spectrum of the Hamiltonian (\ref{Hcylinder}), which depends continuously on the magnetic flux, is

\begin{equation}
\label{spectrum}
\epsilon_{p,q}(\phi)=2t\cos(\frac{2\pi}{N}(p+\phi))+2t\cos(\frac{\pi}{M+1}q)
\end{equation} with $p$ and $q$ two integers such that $-N/2 \leq p \leq N/2-1$ and $ 1 \leq q \leq M$. We have applied open boundary conditions at the ends of the cylinder. Note that the spectrum, and therefore every thermodynamic quantity, is periodic in flux, with periodicity $\phi=1$ \cite{byers}. As the magnetic field is increased, the energy levels evolve and many level crossings appear \cite{pleutin} (Cf. Fig. 1). The details of the  pattern of the crossing points is very complicated and depends on the geometry of the system and parameters of the model.

At zero temperature, the ground state energy of the system with $N_e$ electrons is obtained by filling up successively the lowest energy levels according to the Pauli principle. Here, we restrict our study, almost exclusively, to the case of equal numbers of up and down spins i.e. $S_z=0$, where the lowest $N_e/2$ levels are doubly occupied. The important variations of the energy levels with the magnetic field cause changes in the level occupation. As a consequence of that, at certain values of the magnetic field, $\phi_c$, a former excited state may become the new ground state. Such a switch of ground state are responsible for the appearance of cusps in the ground state energy curve. We can see an example in Fig. 2a for a cylinder with $N=M=10$ and 80 electrons; the ground state energy as function of the magnetic flux shows 5 different cusps. At these particular points, the ground state is changed from a state $|\Psi_i>$ to a new one $|\Psi_o>$ ($i$ for `{\it in}', and $o$ for `{\it out}' to do an analogy with scattering theory) which are both a Slater determinant built with the one-electron states, $\{\varphi_{p,q}\}$, eigenfunctions of the Hamiltonian (1). The two determinants differ only by the highest occupied level, $\varphi^H_{p_i,q_i}$ and $\varphi^H_{p_o,q_o}$ (where the upperscript $H$ is for Highest).

One may see this orbital effect as a succession of scattering events where the time is replaced by the magnetic flux. The system of $N_e$ particles evolves freely until a particular `time', $\phi_c$, where the most energetic particle is scattered: $\varphi^H_{p_i,q_i} \rightarrow \varphi^H_{p_o,q_o}$. Depending on the one-electron state exchanged, one should distinguish between
\begin{itemize}
\item `{\it forward scattering}' (FS) where the two one-electron states, $(p_i,q_i)$ and $(p_o,q_o)$, correspond to particles moving in the same direction along the circumference,
\item `{\it backward scattering}' (BS) where the two one-electron states, $(p_i,q_i)$ and $(p_o,q_o)$, correspond to particles moving in the opposite direction along the circumference.
\end{itemize} In general, a FS corresponds to smaller change in momentum, $\delta k=([(p_i-p_o)\frac{2 \pi}{N}]^2+[(q_i-q_o)\frac{\pi}{M+1}]^2)^{1/2}$, than a BS. In Fig 2a, only the cusp at $\phi \simeq 0.12$ is associated with a FS event, all the others are BS events.

The persistent current (PC) is a thermodynamic quantity given, at zero temperature, in terms of the ground state energy $E$ by

\begin{equation}
\label{pcurrent}
I_{PC}=-\frac{\partial E(\phi)}{\partial\phi}=-2\sum_{(p,q)_{occ}}\frac{\partial \epsilon_{p,q}(\phi)}{\partial\phi}.
\end{equation} The second equality arises for free electrons only and in the case where $S_z=0$. It can be shown that this derivative is proportional to the average of the current operator. This current yields an orbital magnetic moment which can be detected experimentally \cite{expring}. Obviously, the persistent current presents discontinuities for each value of the magnetic flux where the ground state energy shows cusps. Fig. 2b gives an example for the very same cylinder ($N=M=10$, $N_e=80$). Note that the persistent current in mesoscopic cylinders was studied with some details in the past. For instance, it was shown in \cite{szopa2} that its intensity strongly depends on the shape of the Fermi surface. This property has important consequences for carbon nanotubes \cite{szopa}. Most of the other studies insist on the role played by disorder \cite{bouzerar2,roche2}. As a remark, one may say that the orbital effect described above, works to reduce the persistent current: if one would keep the level occupation frozen and then, let evolve the ground state energy as function of the magnetic flux the resulting PC would be more than one order of magnitude larger.

The discontinuities seen in the PC (Fig. 2b) is a general phenomenon, within the free electron picture, which affects any response function. For instance, in ref. \cite{pleutin}, the static electric polarisability of diverse cylindrical systems was studied and shown to present these characteristics. It was then suggested to use this physical quantity to get some insights into the electronic structure of nanoscopic materials such as carbon nanotubes.

In more realistic situations, the neighbouring energy levels (Fig. 1) are coupled by various interactions i.e. Coulomb interaction, disorder. In consequence, the levels will not cross each other but rather will come close and then repel in avoided crossings. Therefore, the different response functions of the system  will not show discontinuities but rather abrupt changes at the position of the avoided crossings. Next we consider the case of long and short range Coulomb potential in the weakly interacting limit assuming the system without disorder.

\section{Interacting electrons. Coulomb and spin effects.}

In this section, we consider the same cylinders pierced by a magnetic flux but, with interacting electrons. Therefore, we consider the following Hamiltonian 

\begin{equation}
\label{fullhamiltonian}
\hat{H}=\hat{H}_0+\hat{H}_{int}
\end{equation}where $\hat{H}_0$ describes free electrons in applied magnetic field and has been defined previously (see Eq. (\ref{Hcylinder})). The second term introduces the Coulomb interaction 

\begin{equation}
\label{coulomb}
\hat{H}_{int}=\frac{1}{2}\sum_{(n,m) \sigma,(n',m')\sigma'}U_{(n,m),(n',m')}c^{\dagger}_{n,m,\sigma}c^{\dagger}_{n',m',\sigma'}c_{n',m',\sigma'}c_{n,m,\sigma}
\end{equation}where $U_{(n,m),(n',m')}$ is assumed to be either short range (Hubbard model) or long range. In the case of long range potential, we choose, to be specific, the Ohno potential known in the chemical literature where it is often used to describe $\pi$ electrons in organic materials such as conjugated polymers or carbon nanotubes

\begin{equation}
\label{ohno}
U_{(n,m),(n',m')}=\frac{U}{\sqrt{1+a_0 r^2_{(n,m),(n',m')}}}.
\end{equation}It includes a kind of effective screening due to the inner electrons via an effective screening length, $a_0$, with typical value of $0.611\AA^{-2}$ \cite{baeriswyl}. $r_{(n,m),(n',m')}$ is the distance between two electrons siting in sites $(n,m)$ and $(n',m')$ given in angstr\"om; we choose the lattice units to be close to the usual carbon-carbon bond length in graphite, $a=1.4\AA$. The functional form of the potential is then fixed and only $U$ is kept as a variable. Apart the specificity of this particular potential, its main characteristics are rather general: a $1/r$ behaviour at large distances and an effective screening which prevents from any discontinuities at short distances. It could then also be taken as a reasonable Coulomb potential for other systems such as semiconducting rings or, for $a_0$ very small (Hubbard model), metallic rings. 

\subsection{Exact diagonalisation studies}

We have first considered very small cylinders for which it is possible to diagonalise exactly the Hamiltonian with interaction (Eq. (\ref{fullhamiltonian})). Hereafter, particular results for a cylinder with $N=3$, $M=2$ and $N_e=4$ (two up and two down electrons) are discussed. For this filling, the ground state energy curve shows one crossing point in magnetic field in the case without interaction. The results are shown in Fig. 3 for increasing values of the Coulomb potential. Three effects of the Coulomb interaction can be seen.

\begin{itemize}
\item First, as expected - but it is not clearly shown in the figure - the crossing point of the free-electron model is replaced by an avoided crossing: the Coulomb interaction opens a gap. Moreover, the avoided crossing becomes more and more pronounced with increasing values of the Coulomb interaction.
\item Second, the position in magnetic field of the (avoided) crossing point is shifted, $\phi_c \rightarrow \tilde{\phi_c}$, and the importance of the shift increases with the intensity of the interaction. In the following, we call this effect the `{\it Coulomb effect}'.
\item Third, for large enough $U$, one notes the formation of a new plateau like structure in the ground state energy curve at the position of the (avoided) crossing point whose size increases with the interaction strength. Because of the appearance of this plateau, two new real crossing points are formed. We will see latter that this plateau may be explained by the spin degree of freedom: this is a `{\it spin effect}'.
\end{itemize}

These effects cause by the Coulomb interaction, will be studied in more details in the following two subsections using simple approximations allowing for the study of larger systems.

\subsection{Two reference ansatz. Coulomb effect.}

We focus mainly on the weak-interacting regime where a mean field theory is supposed to be in principle, a good starting point. However, as we have already seen in the previous section, the free electron ground state becomes degenerate for some particular values of the magnetic field (cf. Fig. 2a). The Coulomb interaction contributes to mix the diverse electronic configurations thereby, at the vicinity of these points of degeneracy, one expects the ground state of the system to be a true many body state unable to be described by only one Slater determinant: a mean field description is then not appropriate. We propose, in this subsection, a simple variational ansatz going beyond a mean field treatment and able to capture some of the important many-body effects.

It was suggested long ago \cite{frenkel}, to extent the usual Hartree-Fock theory for linear combination of Slater determinants

\begin{equation}
\label{multireference}
|\psi>=\sum_k\alpha_k |\Phi_k>
\end{equation}where $\alpha_k$ are variational parameters and $|\Phi_k>$ are some chosen Slater determinants built from one-electron states determined by the variational principle \cite{mrs,levy}. The choice of Slater determinants entering the composition of $|\psi>$ is, of course, motivated by the problem under studies. This kind of theory is particularly relevant in case of degeneracy such that appearing here for some values of the magnetic flux, $\phi_c$. 

Before studying the minimal ansatz of the form (\ref{multireference}) relevant for our particular problem, we start by introducing new operators which are linear combinations of the site operators seen previously

\begin{equation}
\label{operatorA}
A_{i,\sigma}=\sum_{n,m}a^i_{n,m}c_{n,m,\sigma}
\end{equation} where $a^i_{n,m}$ are complex coefficients. At best, these coefficients are determined after a variational procedure as it will be done latter. In the free electron model, these operators correspond to the molecular orbitals, eigenfunctions of $\hat{H}_0$. In any case, they are listed by increasing value of their corresponding energy.

In the non-interacting case, the ground state energy shows points of degeneracy as function of the magnetic flux. Let us first, analyse in details these different crossing points, this will help us to define our ansatz. For $S_z=0$, each of these points shows, at least, a fourfold degeneracy. The degenerate states are listed in the following.

\begin{equation}
\label{ref1}
|\psi_I>=\prod_{i=1}^{\bar{N}}A^{\dagger}_{i\uparrow}A^{\dagger}_{i\downarrow}|0>
\end{equation}where $\bar{N}=N_e/2$ and $|0>$ is the vacuum without electron. This state is the state $|\psi_i>$ mentioned above. The three other degenerate states are
\begin{equation}
\label{ref2}
|\psi_{II}>=A^{\dagger}_{\bar{N}+1\downarrow} A^{\dagger}_{\bar{N}+1\uparrow}A_{\bar{N}\downarrow} A_{\bar{N}\uparrow}|\psi_I>
\end{equation}this is the state $|\psi_o>$,

\begin{equation}
\label{ref3}
|\psi_{III}>=A^{\dagger}_{\bar{N}+1\uparrow} A_{\bar{N}\uparrow}|\psi_I>
\end{equation}and

\begin{equation}
\label{ref4}
|\psi_{IV}>=A^{\dagger}_{\bar{N}+1\downarrow} A_{\bar{N}\downarrow}|\psi_I>.
\end{equation} This analysis is illustrated in Fig. 4 where the four Slater determinants are represented schematically together with the behaviour of their corresponding energies as function of the magnetic field.

The electronic configurations $|\psi_I>$ and $|\psi_{II}>$ are closed-shell. With Coulomb interaction, they are expected to give the main components of the ground state at values of the magnetic flux sufficiently lower or higher than $\phi_c$, respectively. The two other configurations, $|\psi_{III}>$ and $|\psi_{IV}>$, are mono-excitations with respect to $|\psi_I>$ or $|\psi_{II}>$. They are expected to play a role in the direct vicinity of $\phi_c$ (Fig. 4). As a first trial, we consider only the closed-shell determinants in the linear combination (\ref{multireference}); we will see, at the end, that this simple ansatz gives good results not too close to the degeneracies and is sufficient to describe (i) the avoided crossing formation  and (ii) the Coulomb effect but not the spin effect. Within this approximation, we consider the following two-reference ansatz

\begin{equation}
\label{ansatz}
|\psi>=\alpha|\psi_I>+\beta|\psi_{II}>=[\alpha\hat{1}+\beta A^{\dagger}_{\bar{N}+1\downarrow} A^{\dagger}_{\bar{N}+1\uparrow}A_{\bar{N}\downarrow} A_{\bar{N}\uparrow}]\prod_{i=1}^{\bar{N}}A^{\dagger}_{i\uparrow}A^{\dagger}_{i\downarrow}|0>
\end{equation}where $\hat{1}$ is the unit operator. The expansion coefficients, $\alpha$ and $\beta$, and the set of coefficients $\{a^i_{n,m}\}$ used to define the one-body wave functions (\ref{operatorA}) are now determined from the variational principle. More precisely, we look for an extremum of the ground state energy:

\begin{equation}
\label{egs}
{\cal E}(\alpha,\beta,\{a^i_{n,m}\})=\frac{<\psi|\hat{H}|\psi>}{<\psi|\psi>}.
\end{equation}

We use a two-step iterative procedure to determine the best wave-function $|\psi>$. In a first step, the expansion coefficients $\alpha$ and $\beta$ are determined by diagonalisation of the following two by two matrix

\begin{equation}
\label{ci}
\left(\begin{array}{c}<\psi_I|\hat{H}|\psi_I> \quad \quad <\psi_I|\hat{H}|\psi_{II}>\\
<\psi_{II}|\hat{H}|\psi_I> \quad \quad <\psi_{II}|\hat{H}|\psi_{II}>\end{array}\right).
\end{equation}In a second step, the coefficients ${a^i_{n,m}}$ are determined keeping the expansion coefficients, $\alpha$ and $\beta$ fixed. Minimisation of ${\cal E}(\alpha,\beta,\{a^i_{n,m}\})$ with respect to the set $\{a^i_{n,m}\}$ gives the so-called generalised Brillouin theorem for multiconfigurational Hartree-Fock theory \cite{mrs,levy}, valid for any multireference states (\ref{multireference})

\begin{equation}
\label{brillouin}
<\psi|[A^{\dagger}_rA_s,\hat{H}]|\psi> = 0.
\end{equation} The Brillouin theorem for usual HF theory states that the matrix elements of the total Hamiltonian between the ground state and any mono-excitations cancel out. Its generalisation to multireference HF theory is less clear: it means that the matrix elements of the total Hamiltonian between the ground state and some linear combinations of excited configurations cancel out. Using this condition, it is possible to calculate the set of coefficients $\{a^i_{n,m}\}$ as we will see below. Then, the process is continued by repeating these two steps until convergence is reached.

The first step is straightforward. For the second step, to proceed starting from Eq. (\ref{brillouin}), we follow closely the Ref. \cite{mrs}.  In order to make use of the generalised Brillouin theorem, it is convenient to rewrite the Hamiltonian as

\begin{equation}
\label{ham}
\hat{H}=\hat{F}-\hat{V}+\hat{H}_{int}
\end{equation}where $\hat{V}$ is an effective one-body operator and, $\hat{F}$ is the Fock operator which we require to be diagonal

\begin{equation}
\label{fock}
\hat{F}=\hat{H}_0+\hat{V}=\sum_{r,s,\sigma}\delta_{rs}\epsilon_r A^{\dagger}_{r,\sigma}A_{r,\sigma}.
\end{equation} With this rewriting of the Hamiltonian, the generalised Brillouin theorem is used to determine $\hat{V}$ and, thereby, to build the Fock operator. By using a multireference ansatz such as (\ref{multireference}) and (\ref{ansatz}), the Fock operator takes a block structure

\begin{equation}
\label{fockblock}
F=\left(\begin{array}{c} F_{cc} \quad F_{co} \quad F_{ce}\\
F_{co} \quad F_{oo} \quad F_{oe}\\
F_{ce} \quad F_{oe} \quad F_{ee}\end{array}\right).
\end{equation}Let us name $\xi_r\chi_{\sigma}$, with $1 \le r \le N.M$ and $\sigma=\pm1/2$, the variational one-electron wave functions associated with the fermionic operators $A^{\dagger}_{r,\sigma}$ and $A_{r,\sigma}$. $\xi_r$ is the orbital part and $\chi_{\sigma}$ the spin part. They are listed by increasing values of energy. The one-electron operator $\hat{F}_{cc}$ is defined in the subspace spanned by $\xi_r$ with $1 \le r \le \bar{N}-1$ ($\bar{N}=N_e/2$), $\hat{F}_{oo}$ in the subspace spanned by $\xi_{\bar{N}}$ and $\xi_{\bar{N}+1}$ and $\hat{F}_{ee}$ in the subspace spanned by $\xi_r$ with $r> \bar{N}+1$. It is important to stress that the effective Fock operator is non-uniquely defined by the generalised Brillouin condition. Indeed, only the off-diagonal blocks are determined by using Eqs. (\ref{brillouin}) and (\ref{fock}). Additional assumptions are needed to fix the three diagonal blocks: here, we calculate these blocks as matrix elements of the one-particle operator $\hat{F}_{ce}$, procedure particularly appropriate for closed shell configurations \cite{mrs}.

In order to test the quality of the variational wave function (\ref{ansatz}), we have first considered very small cylinders and compare the approximate results with the results obtained by exact diagonalisation. Fig. 5 shows a comparison between the variational calculation and the exact result for weak interaction ($U=0.1t$ and $U=0.2t$). A good agreement is obtained which becomes excellent away from the crossing points. Indeed, Fig. 5 shows only the direct vicinity of the crossing point. As we can see, our simple variational ansatz (\ref{ansatz}) describes well the Coulomb effect i.e. the shift of the crossing point, but is less accurate at the crossing itself. The reasons for that could be the spin effect and/or the two open-shell Slater determinants not included in the ansatz (\ref{ansatz}). 

Before going further, let us discussed in some details the origin of the Coulomb effect well described by our ansatz. For simplicity, we neglect the off-diagonal term of the two by two matrix Eq. (\ref{ci}) - responsible for the opening of the gaps - and the Fock term keeping only the most important contribution of the Coulomb interaction, namely the Hartree term. As a last approximation, instead of considering the self-consistent procedure just described we keep only the first order correction. Without interaction, at every cusp, $\phi_c$, the system changes of ground state from $|\Psi_i>$ to $|\Psi_o>$. The energies of these two states are $E_i(\phi)$ and $E_o(\phi)$. They are both two parabola-like curves that cross at $\phi_c$, $E_i(\phi_c)=E_o(\phi_c)$. The electron densities of these two states are $\rho_i(n,m,\phi)$ and $\rho_o(n,m,\phi)$, respectively, which differ in general i.e. $\rho_i(n,m,\phi) \ne \rho_o(n,m,\phi)$,

\begin{equation}
\label{density}
\rho_{\alpha}(n,m,\phi)=2\sum_{(p_{\alpha},q_{\alpha})_{occ.}}|\varphi_{p_{\alpha},q_{\alpha}}(n,m,\phi)|^2 
\end{equation}where $\alpha=i/o$. The first order Hartree corrections, $W_i(\phi)$ and $W_o(\phi)$, are given by

\begin{equation}
\label{hartree}
W_{\alpha}(\phi)=\frac{1}{2}\sum_{(n,m)\atop (n',m')}\rho_{\alpha}(n,m,\phi) U_{(n,m),(n',m')}\rho_{\alpha}(n',m',\phi)\left(1-\frac{1}{2}\delta_{(n,m),(n',m')} \right).
\end{equation} Since the densities differ, the magnitude of the Hartree corrections will also be different - in particular, at the crossing point, $W_i(\phi_c) \ne W_o(\phi_c)$). In consequence of that, the two parabola, $E_i(\phi)$ and $E_o(\phi)$, are shifted more or less strongly with Coulomb interaction to higher energies by the Hartree corrections, $W_i(\phi)$ and $W_o(\phi)$. Since the Hartree terms have different magnitude, the two-shifted parabola will cross at a different value of the magnetic flux, $\tilde{\phi_c}$, such that $E_i(\tilde{\phi_c})+W_i(\tilde{\phi_c}) = E_o(\tilde{\phi_c})+W_o(\tilde{\phi_c})$. This is the Coulomb effect illustrated schematically on Fig. 6 and observed, for instance, in the exact results shown in Fig 3 and Fig. 5.

The magnitude of the shift is determined mainly by the difference between the Hartree contributions, $\delta W(\phi)=W_i(\phi)-W_o(\phi)$, caused by the highest occupied states, $\varphi^H_{{\bf p_i,q_i}}$ and $\varphi^H_{{\bf p_o,q_o}}$. If one uses more compact notation replacing the two pairs of indices, $(p_i,q_i)$ and $(p_o,q_o)$, by ${\bf k_i}$ and ${\bf k_o}$, respectively, and if one works in the momentum space instead of the real space, the mean contribution to $\delta W(\phi)$ is given, at first order in perturbation theory, by the following equation
\begin{equation}
\label{hartreedifference}
\delta W(\phi) \propto \sum_q (|\varphi^H_{{\bf k_i}}|^2U_{\bf{q}}|\varphi_{{\bf k_i+q}}|^2-|\varphi^H_{{\bf k_o}}|^2U_{\bf{q}}|\varphi_{{\bf k_o+q}}|^2) \sim \sum_q(|\varphi^H_{{\bf k_i}}|^2U_{\bf{q}}|\varphi_{{\bf k_i+q}}|^2-|\varphi^H_{{\bf k_o}}|^2U_{\bf{q+\delta q}}|\varphi_{{\bf k_i+q}}|^2)
\end{equation}where the summation is done for states that are occupied in the ground state, $U_q$ is the Fourier transform of the Coulomb potential and ${\bf\delta q}={\bf k_0}-{\bf k_i}$. On the basis of this approximate formula, one can draw some conclusions. First, one sees that the magnitude of $\delta W(\phi)$ would be enhanced by a long range Coulomb potential which gives diverging contribution at small momenta, $q$, (for instance, the Coulomb interaction behaves as $1/q$ for a two dimensional system) on the contrary to short range interaction that provides only constant contribution (independent of $q$). Second, If $\delta q$ is small, as it happens for Forward Scattering (FS), one may neglect in the formula (\ref{hartreedifference}) the difference between the one-particle states $\varphi^H_{{\bf k_i}}$ and $\varphi^H_{{\bf k_o}}$. Then, $\delta W(\phi)$ becomes roughly proportional to the derivative of the Coulomb potential, $dU_q/dq$, which is a diverging quantity at small values of $q$ (for infinite lattice) in the case of long range potential. To conclude, (i) we expect a large Coulomb effect for long range potential and, comparatively, no effect for short range potential, (ii) in the case of long range potential, since the difference between ${\bf k_i}$ and ${\bf k_o}$ are smaller for FS crossing points, than BS crossing points, the Coulomb effect are expected to be stronger for FS than BS. These two conclusions are confirmed both by exact diagonalisation for small cylinders (see Figs 3, 11 and 12) and by using our variational ansatz as we will see below.

We consider a bigger cylinder with $N=10$, $M=10$ and $N_e=80$ ($40$ electrons up and $40$ electrons down). Then, it is not possible anymore to do exact calculation and one has to rely on the approximate result. Important shifts of all the crossing points are obtained (cf. Fig. 7). As it is expected from the qualitative argument above, the most important shift is obtained for FS type of crossing points: in our example a shift of more than $5.10^{-2}\phi_0$ is obtained which could correspond to significant value of magnetic field. Last, note that with on-site interaction only, no differences compared to the non-interacting case can be detected at the scale of the figure. This confirms that the magnitude of the shifts is controlled by the range of the Coulomb potential, in agreement with our qualitative argument based on Eq. (\ref{hartreedifference}). The screening of the Coulomb potential, induced by the proximity of a metallic electrode, for instance, will tend to decrease the magnitude of the Coulomb effect. 

The corresponding persistent current is shown in Fig. 8 compared to the one obtained for the very same cylinder but without Coulomb interaction. The discontinuities appear shifted to higher magnetic flux due to the Coulomb effect. This is particularly sensible for the FS event for which, moreover, the magnitude of the discontinuity is enhanced.

\subsection{Spin effect}

It is clear from the results of exact diagonalisation (cf. Fig. 3), that the ansatz (\ref{ansatz}) is not sufficient for relatively high Coulomb interaction. According to the analysis of the crossing points, we have missed in the linear combination (\ref{multireference}), at least, the two mono-excitations $|\psi_{III}>$ and $|\psi_{IV}>$. In the non-interacting case ($U=0$), these two configurations are degenerate in energy, but, in the interacting case ($U \ne 0$), they are mixed into two components that split from each other: a singlet state, at high energy, and a triplet state, at low energy, with a gap of $2\Gamma$, $\Gamma$ being the exchange energy

\begin{equation}
\label{exchange}
\Gamma=|<\psi_{III}|\hat{H}|\psi_{IV}>|.
\end{equation}The corresponding wave functions are given by the symmetric and antisymmetric linear combinations

\begin{equation}
\label{triplet}
|\psi_{T/S}>=\frac{1}{\sqrt{2}}[|\psi_{III}>\mp|\psi_{IV}>]=\frac{1}{\sqrt{2}}[A^{\dagger}_{N+1,\uparrow}A_{N,\uparrow}\mp A^{\dagger}_{N+1,\downarrow}A_{N,\downarrow}]|\psi_I>
\end{equation} where the indices $T$ is for {\it Triplet} and $S$ for {\it Singlet}. The triplet state may be lower in energy than the variational ground state found in the previous subsection. 

In principle, a first improvement of the ansatz (\ref{ansatz}) would consist to add the components (\ref{ref3}) and (\ref{ref4}) to the trial wave function. This would increase the number of variational parameters by two and, more importantly, since $|\psi_{III}>$ and $|\psi_{IV}>$ are both mono-excitations i.e. open shell configurations, this would force us to change the procedure of calculation. To give an analogy with usual HF theory, one would have to change from a restricted Hartree-Fock to unrestricted Hartree-Fock type of calculation \cite{mrs}. We postpone this treatment to forthcoming work and, instead, adopt a kind of minimal strategy to get a first understanding of  what could be the role of the spin at the vicinity of the crossing points. Starting from the variational solution (\ref{ansatz}), we build the triplet wave function (\ref{triplet}) and then estimate its energy
\begin{equation}
\label{energieT}
{\cal E}_T(\alpha,\beta,\{a^i_{n,m}\})=<\psi_T|\hat{H}|\psi_T>
\end{equation}keeping the parameters $\alpha$, $\beta$ and $\{a^i_{n,m}\}$ frozen. The results obtained for the triplet state are then not variational, but should give, nevertheless, useful indications about the total spin of the ground state as function of the applied magnetic field. At the vicinity of the crossing points, the triplet state can be lower in energy suggesting a possible spin transition induced by a longitudinal magnetic field. Such an example is shown in Fig. 9 for a very small cylinder with $U=t$. The two reference ansatz (\ref{ansatz}) reproduces well the shifting of the crossing point (see also Fig. 5), but the additional `plateau' induced by the interaction is missed with this approximation. However, it is partially described if one considers also the energy (\ref{energieT}): on a certain interval of magnetic flux - corresponding to this plateau - the triplet state (\ref{triplet}) is lower in energy. Our procedure is in a good agreement with the exact results and, therefore, strongly suggests the following sequence for the total electronic spin at the vicinity of each point of (near) degeneracy and for increasing values of magnetic field, $S=0\rightarrow S=1 \rightarrow S=0$. At each spin transition appears a new real crossing points. Fig. 10 shows the corresponding persistent current compared with the one obtained for the non-interacting case. The long-range Coulomb interaction shifts the discontinuities - this is the Coulomb effect - and is responsible for the appearance of a new `plateau' - the {\it Triplet} plateau - which may be detected experimentally.

The above result may be explained by an usual qualitative argument based on the Hund's rule. In case of double degeneracy, or near degeneracy, such as encountered at the direct vicinity of each crossing point, and for $S_z=0$, the Coulomb interaction will favour the state where the two degenerate levels, $\varphi^H_{p_i,q_i}$ and $\varphi^H_{p_o,q_o}$, are mono-occupied. Indeed, it costs more Coulomb energy to doubly occupied one of these two levels. Moreover, the repulsive interaction will favour the state with maximum total spin, here $S=1$, since the coordinate wave function is then anti-symmetrised which allows to gain the exchange energy. The two highest electrons will occupy the levels $\varphi^H_{p_i,q_i}$ and $\varphi^H_{p_o,q_o}$ with total spin one, as long as the energy difference between the levels remains smaller than the energy gain due to Coulomb interaction. At this point, it is important to remember that we have not considered the Zeeman interaction which may change the picture at high magnetic field.

Before doing the same analysis for a bigger cylinder, we present some exact diagonalisation studies of the very same small cylinder but with a short-range Coulomb potential i.e. namely by using the Hubbard model. The results are summarised in Fig. 11 where the ground state energy is shown as function of the magnetic flux for several values of $U$. These results have to be compared with the ones of Fig. 3 presenting the same quantity but using the long-range Coulomb potential (Eq. \ref{ohno}). One sees, first, that the Coulomb effect is considerably reduced by the use of an on-site term only: there is almost no shifting of the crossing points. Second, the spin effect is, on the contrary, sensibly enhanced in the sense that (i) the {\it Triplet} plateau appears for lower values of $U$ than in the case of long range Coulomb potential and, (ii) this plateau can be much more extended in magnetic flux. It seems that there is a kind of competition between the two effects. The more the Coulomb effect is important the more the size of the {\it Triplet} plateau is reduced. This can be understood by the following qualitative argument.

The difference in energy between the variational ground state and the Triplet state is given by
\begin{equation}
\label{diffgstriplet}\begin{array}{l}
{\cal F}(\alpha,\beta,\{a^i_{n,m}\})={\cal E}(\alpha,\beta,\{a^i_{n,m}\})-{\cal E}_T(\alpha,\beta,\{a^i_{n,m}\})=\\
\\
\quad \quad \quad \frac{1}{4}\sum_{(n,m),(n',m')}(\rho_{i}(n,m)-\rho_{o}(n,m))U_{(n,m),(n',m')}(\rho_{i}(n',m')-\rho_{o}(n',m')).
\end{array}
\end{equation}If ${\cal F}>0$, the Triplet state is the ground state of the system. The on-site part of the Coulomb potential gives always a positive contribution to ${\cal F}$ in favour of a Triplet ground state. On the contrary, since the difference in density, $\delta \rho(n,m)=\rho_{i}(n,m)-\rho_{o}(n,m)$, is in general an oscillatory function, the long range part of the Coulomb potential could give a negative contribution to ${\cal F}$. Therefore, as it is seen in exact diagonalisation studies, the spin effect is expected to be more pronounced for screened interaction.

Next, we have considered a bigger cylinder with $N=10$, $M=10$, $N_e=80$ and $S_z=0$, and evaluated the variational energy (Eq. (\ref{egs})) and the corresponding triplet energy (Eq. (\ref{energieT})). For the long range Ohno potential (Eq. (\ref{ohno})), there is no range of magnetic flux where the {\it Triplet} state would be significantly favoured within our approximation - at least, for relatively low values of $U$ (up to $U=2|t|$ approximately). On the contrary, for the Hubbard potential, there is a $Singlet \rightarrow Triplet$ transition at every crossing points for strong enough $U$. This is shown in Fig. 12 for $U=|t|$. One can notice, a large {\it Triplet} plateau for values of the magnetic flux corresponding to the FS point. 

Last, we have considered exclusively the case with equal numbers of up and down electrons up to now i.e. $S_z=0$. The conclusions will be different with an unpaired electron i.e. $S_z=\pm1/2$. Indeed, in this case, at each crossing point the ground state of the free electron model is only twofold degenerate and not fourfold as it is the case for $S_z=0$. The spin is then not expected to play any role. Therefore, with Coulomb interaction, the avoided crossing and Coulomb effect phenomena should persist but, on the contrary, the spin effect described above should not. We have down exact calculations for very small cylinders. Fig. 13 shows examples for $N=3$, $M=2$, $U=100|t|$, 2 spin up and 1 spin down compared to the case with 4 electrons and $S_z=0$. One clearly sees that, for both long and short range Coulomb potential, the spin effect doesn't occur in the case with one unpaired electron. 

As a remark, one may say that the spin transition found at the vicinity of the crossing points reminds the fractional Persistent Current analysed for purely one-dimensional rings \cite{kusmartsev}. Without interaction, the ground state energy of these systems is periodic with periodicity $\phi_0$ \cite{pleutin}. With infinite interaction, the spin-charge separation phenomenon occurs and the spin excitations of the system contribute to change drastically the behaviour of the ground state energy by changing its periodicity to $\phi_0/N_e$ where $N_e$ is the number of electrons. It would be, of course, very interesting to be able to extrapolate what could happen for the multichannel systems studied here but in presence of very large interaction.

Before the conclusion, one may add that similar spin sequences, as the one pointed out in this subsection, were already predicted to occur in short armchair carbon nanotubes under the influence of an inhomogeneous gate potential \cite{oreg}. The carbon nanotubes have two families of levels which react differently to this applied potential, creating level crossings. Then, for even number of electrons, in the very same way as the situation described here, the first Hund's rule drives the system from a singlet ($S=0$) to a triplet state ($S=1$) at the vicinity of the crossing points. The same phenomenon was also described to happen in quantum dots where the signature of the spin transition was expected to be seen in the variation of the Coulomb-blockade peak positions \cite{ullmo}. 

\section{discussion and conclusion}

We have considered in this work electrons moving on the surface of nanoscopic cylinders described as rolled square lattices. Without disorder - or with a weak disorder - a static magnetic field applied along the cylindrical axis induces points of degeneracy - or quasi-degeneracy. Therefore, with Coulomb interaction, the ground state of the system becomes, at the vicinity of these points, fundamentally a many body state unable to be described by any mean field theory based on only one Slater determinant (as usual Hartree-Fock theory). To circumvent this difficulty, we have studied a simple variational ansatz which is a linear combination of the two closed shell Slater determinants that become degenerate. On top of that, we have also considered the lowest triplet state built out of the variational solution. As a result, we have found three effects induced by the electron-electron interaction at the vicinity of these points of degeneracy.

\begin{itemize}
\item {\it Avoided crossings}. The Coulomb potential induces couplings between the degenerate configurations. As a result, a gap is opened.
\item {\it Coulomb effect}. With repulsive interaction the diverse energy curves, corresponding to diverse Slater determinants, as function of the magnetic flux, are shifted up by the Hartree contributions. Because the magnitude of these contributions differs from configuration to configuration, the Coulomb interaction shifts the position in magnetic field of the (avoided) crossings.   
\item {\it Spin effect}. When degeneracy occurs and for $S_z=0$, the Hund's rule can drive the system to a triplet state. Therefore, at the vicinity of the degeneracy points, the Coulomb interaction can induce a {\it Singlet} $\rightarrow$ {\it Triplet} transition follows by a {\it Triplet} $\rightarrow$ {\it Singlet} one. In the case of one unpaired electron, $S_z=\pm1/2$, the spin effect doesn't occur.
\end{itemize}The Coulomb effect is due mainly to the long range part of the Coulomb potential. On the contrary, the spin effect is favoured by short range interaction. Therefore, a screening of the repulsive interaction induced, for instance, by a gate electrode could change the electronic properties of the system.

The electron-electron interaction, on the one hand, mixes the different configurations and replaces the crossing points of a free electron theory by avoided crossings but, on the other hand, the total spin of the system may be changed due to the Hund's rule creating new real crossing points. Thereby, any response function, providing that the electronic structure of the cylinder is preserved, are expected to show an abrupt change at the position of these crossing points where the total spin is changed. This should be the case in measures of persistent current \cite{expring}, but also, in the static electric magnetopolarisability studied in \cite{pleutin} and already measured for an ensemble of metallic rings \cite{deblock} and, as a last example, in magnetoconductance measurements, with bad contacts to the electrodes, such as the one done in \cite{mwntqd} where multiwall carbon nanotubes behave as quantum dots. Similar effects were studied in \cite{ullmo} for quantum dots, where the spin transitions were shown to give characteristic signatures in the Coulomb-blockade peak positions.

\begin{acknowledgements}
I would like to acknowledge, once more, Prof. Alexander Anatol'evich Ovchinnikov who passed away too early, in March of this year. He introduced me into the field of mesoscopic physics and then continuously provided remarkable insights to push our work ahead. The present paper strongly benefits from his very personal views; they were always deep and precious and they will certainly influence my future work. I immensely miss his sense of humour, his extreme kindless, his numerous advices and, of course, his wonderful talent and ability as a great theoretician.
\end{acknowledgements}

\begin{figure}
\centerline{\psfig{figure=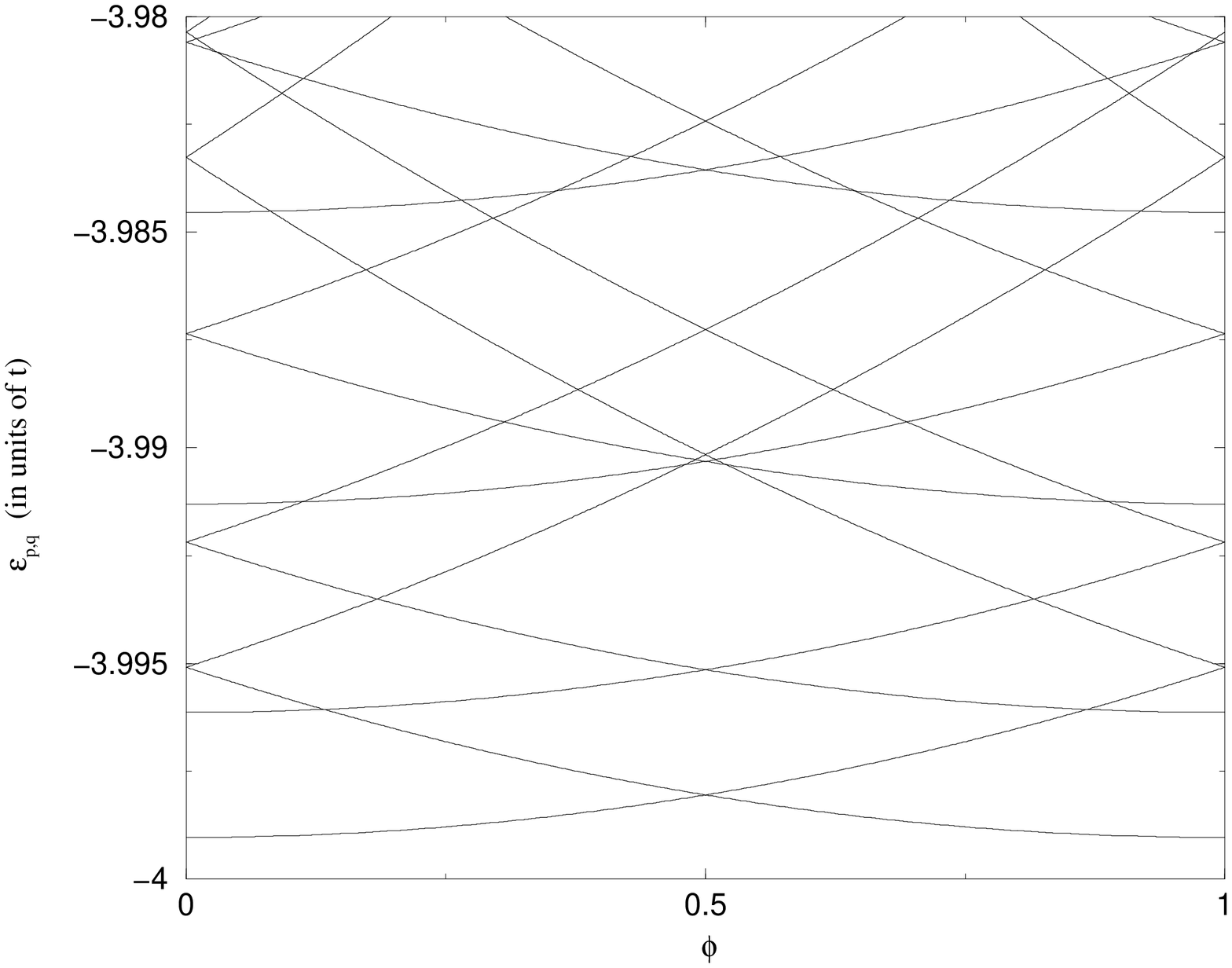,width=15cm,angle=-90}}
\caption{Lowest energy levels of a cylinder with $N=100$ (number of sites along the circumference) and $M=100$ (number of sites along the cylindrical axis) as function of the magnetic flux $\phi$. Numerous level crossings appear at values of $\phi$ which depend on the cylinder geometry.}
\end{figure}

\begin{figure}
\centerline{\psfig{figure=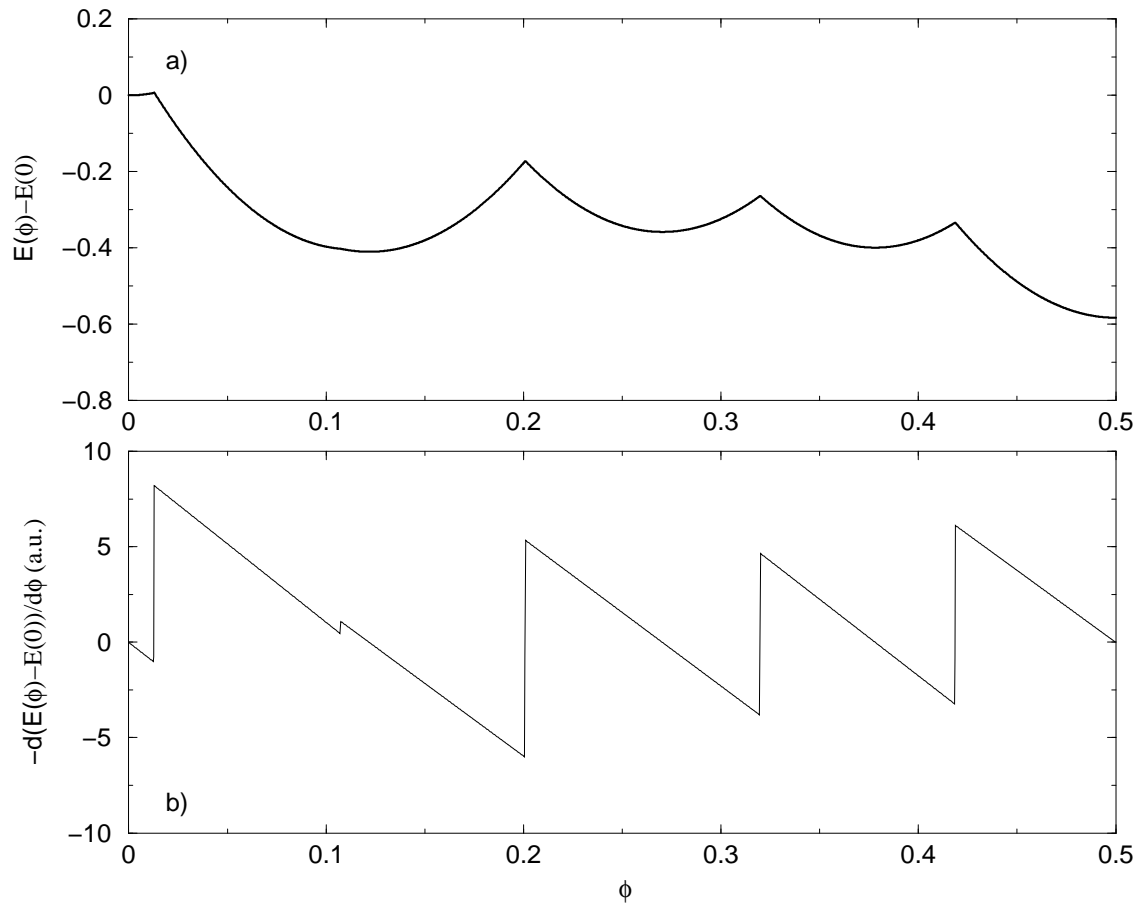,width=15cm,angle=-90}}
\caption{Cylinder with $N=M=10$, $N_e=80$ and $S_z=0$. (a) Energy of the ground state as function of the magnetic flux. Doing an analogy with scattering events, the different cusps are classified into Forward (FS) and Backward (BS) Scattering types (see text). Here, only the cusp at $\phi \simeq 0.12$ is a FS, the other ones are all BSs. (b) The corresponding persistent current. The discontinuities are more pronounced for BS than FS.}
\end{figure}

\begin{figure}
\centerline{\psfig{figure=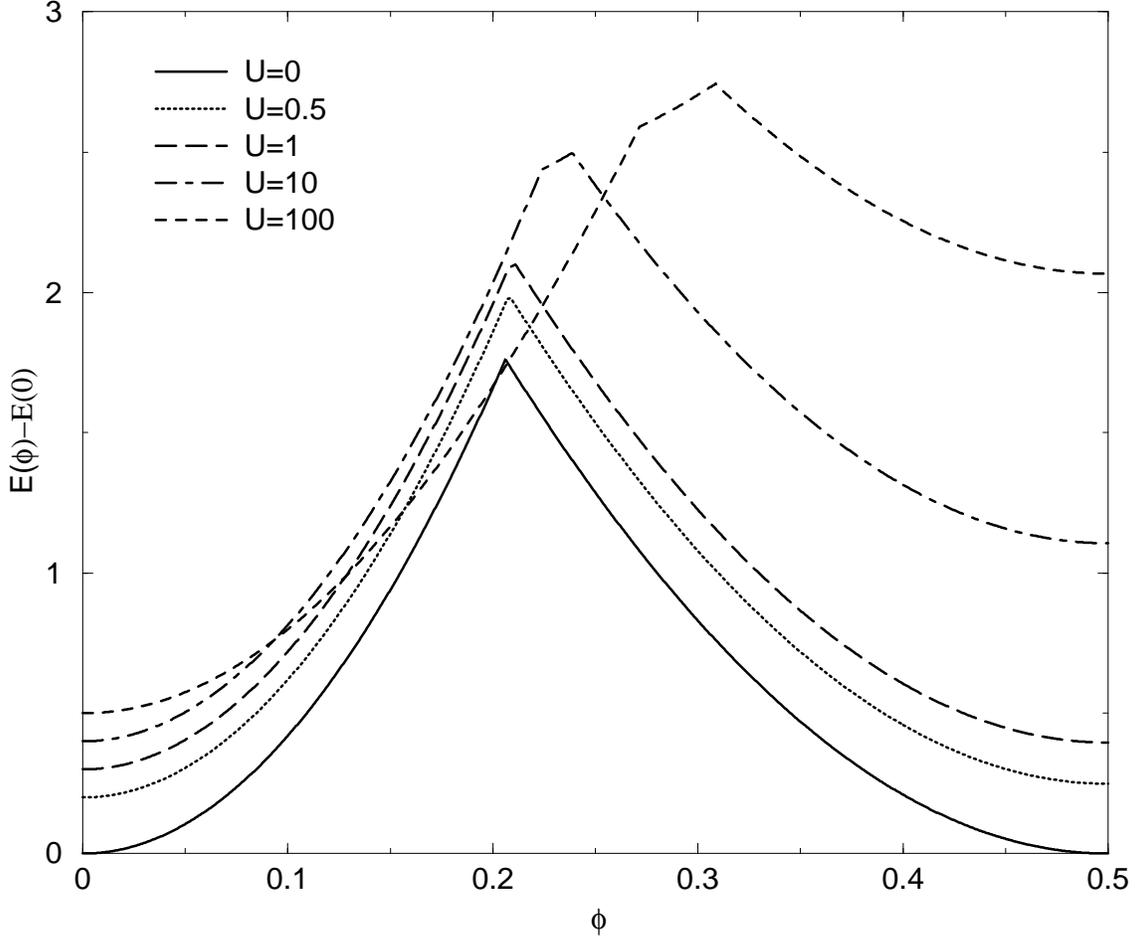,width=15cm,angle=-90}}
\caption{Exact ground state energy as function of the magnetic flux of a small cylinder ($N=3$, $M=2$, $N_e=4$ and $S_z=0$) for increasing values of the long-range Coulomb potential, $U$, in units of $|t|$. The crossing point is shifted to higher magnetic flux with increasing interaction. For strong enough interaction, there is appearance of a new plateau that may be explained by spin excitations. The curves are shifted up for clarity}
\end{figure}

\begin{figure}
\centerline{\psfig{figure=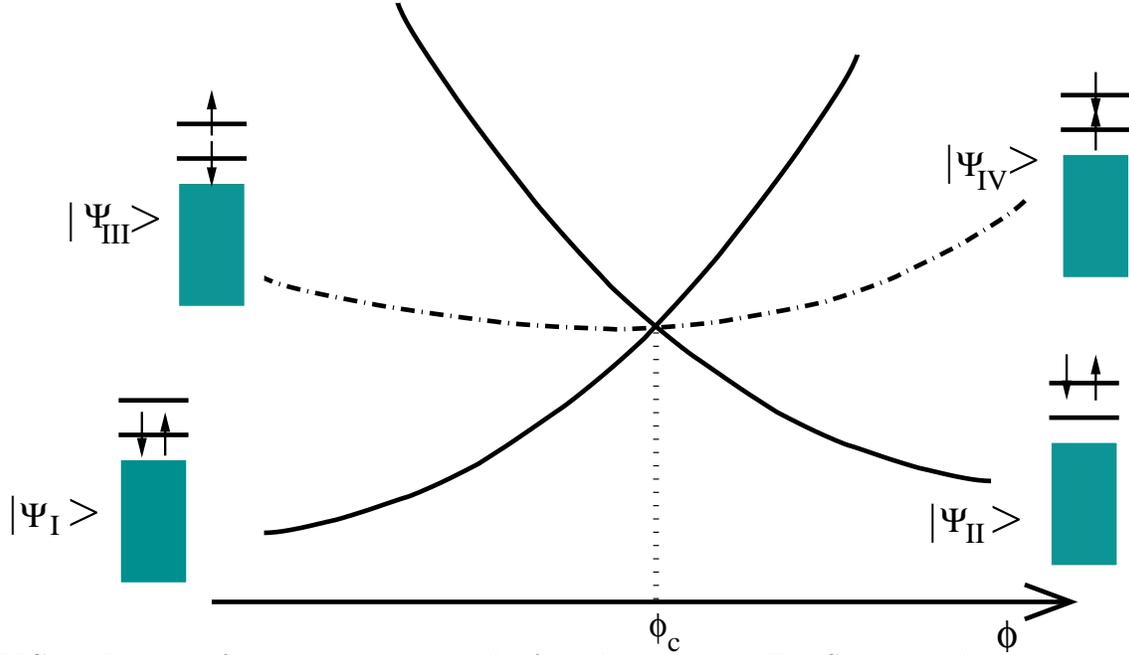,width=15cm,angle=0}}
\caption{Picture of a crossing point in the free electron case. For $S_z=0$,
each crossing point is fourfold degenerate. The four Slater determinants,
$|\psi_I>$, $|\psi_{II}>$, $|\psi_{III}>$ and $|\psi_{IV}>$ (see text) are
shown schematically together with the behaviour of their corresponding energies
as function of the magnetic flux. The full rectangles represent all the doubly
occupied levels except the two highest ones, $\varphi^H_{p_i,q_i}$ and
$\varphi^H_{p_o,q_o}$, which are explicitely shown. In the representation of
the Slater determinants, the energy levels are kept fixed at a particular
value of $\phi$, for simplicity.}
\end{figure}

\begin{figure}
\centerline{\psfig{figure=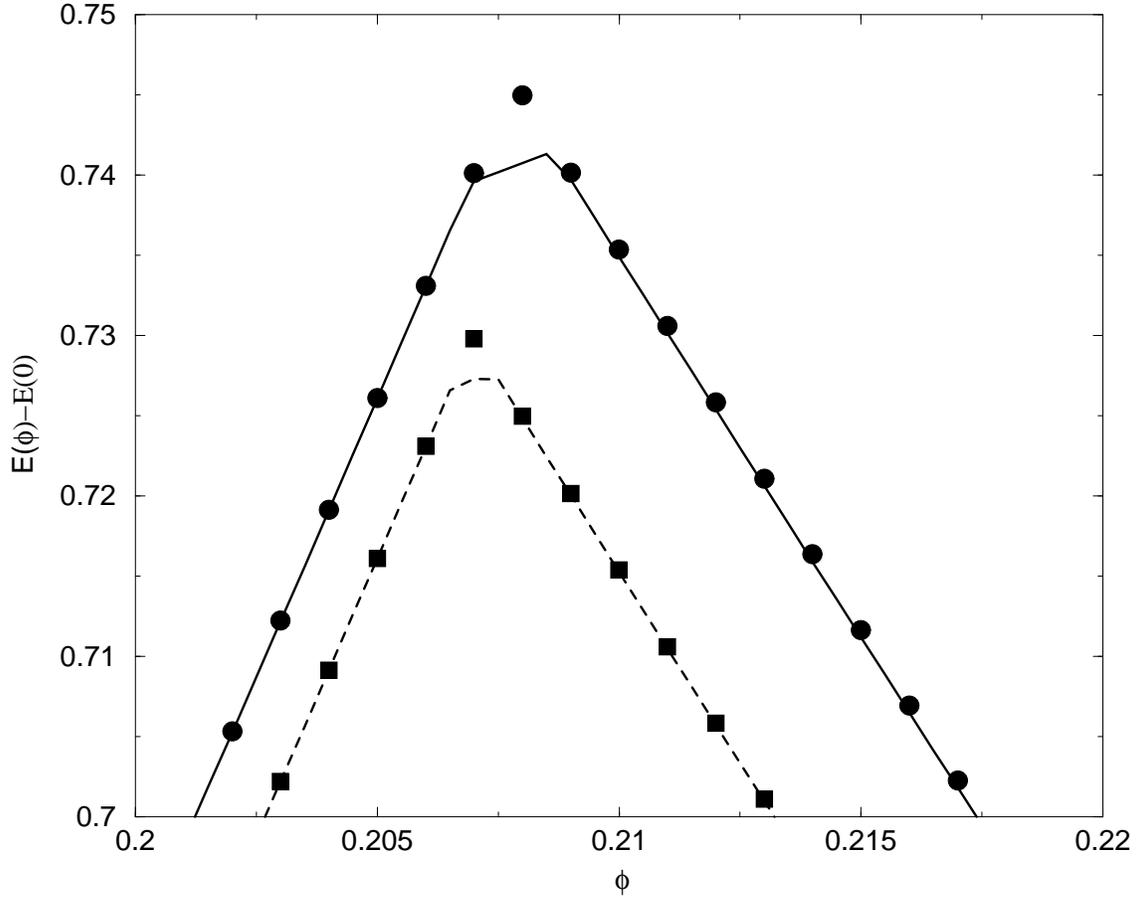,width=15cm,angle=-90}}
\caption{Zoom of the ground state energy of a small cylinder ($N=3$, $M=2$, $N_e=4$ and $S_z=0$) with long range Coulomb interaction in the direct vicinity of the crossing point. The full curve and the dashed curve are the exact results for $U=0.2$ and $U=0.1$ (in units of $|t|$), respectively. The full dots and the full squares are the results obtained with the two-reference variational ansatz (Eq. (\ref{ansatz})) for $U=0.2$ and $U=0.1$ respectively: the Coulomb effect is well reproduced but not the appearance of a plateau which is the result of spin excitations not contained in the ansatz. The curves corresponding to $U=0.1$ are shifted down for clarity.}
\end{figure}

\begin{figure}
\centerline{\psfig{figure=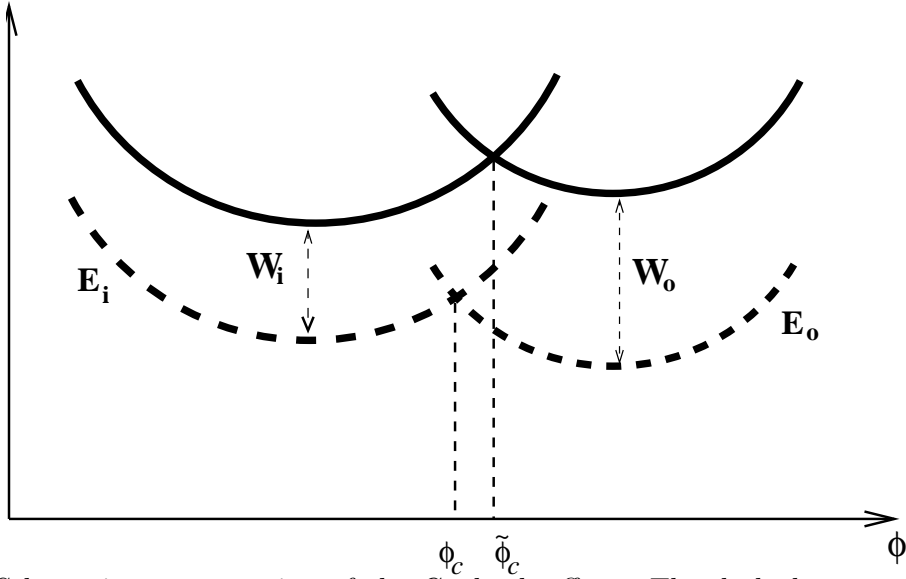,width=12cm,angle=0}}
\caption{Schematic representation of the Coulomb effect. The dashed curves represent the ground state energy of the system without Coulomb interaction: there is a cusp at $\phi_c$ corresponding to a change of ground state $|\psi_i> \rightarrow |\psi_o>$; at this point the ground state is degenerate, $E_i(\phi_c)=E_o(\phi_c)$. The full curves represent the ground state energy with Coulomb interaction: the two initial parabola are shifted up by the Hartree contributions, $W_i$ and $W_o$, which have different amplitudes; consequently, the two interacting ground states are exchanged at a different value of the magnetic flux, $\tilde{\phi}_c$, such that $E_i(\tilde{\phi}_c)+W_i(\tilde{\phi}_c)=E_o(\tilde{\phi}_c)+W_o(\tilde{\phi}_c)$. In other words, the long range Coulomb interaction produces a shift of the cusp, $\phi_c \rightarrow \tilde{\phi}_c$. }
\end{figure}

\begin{figure}
\centerline{\psfig{figure=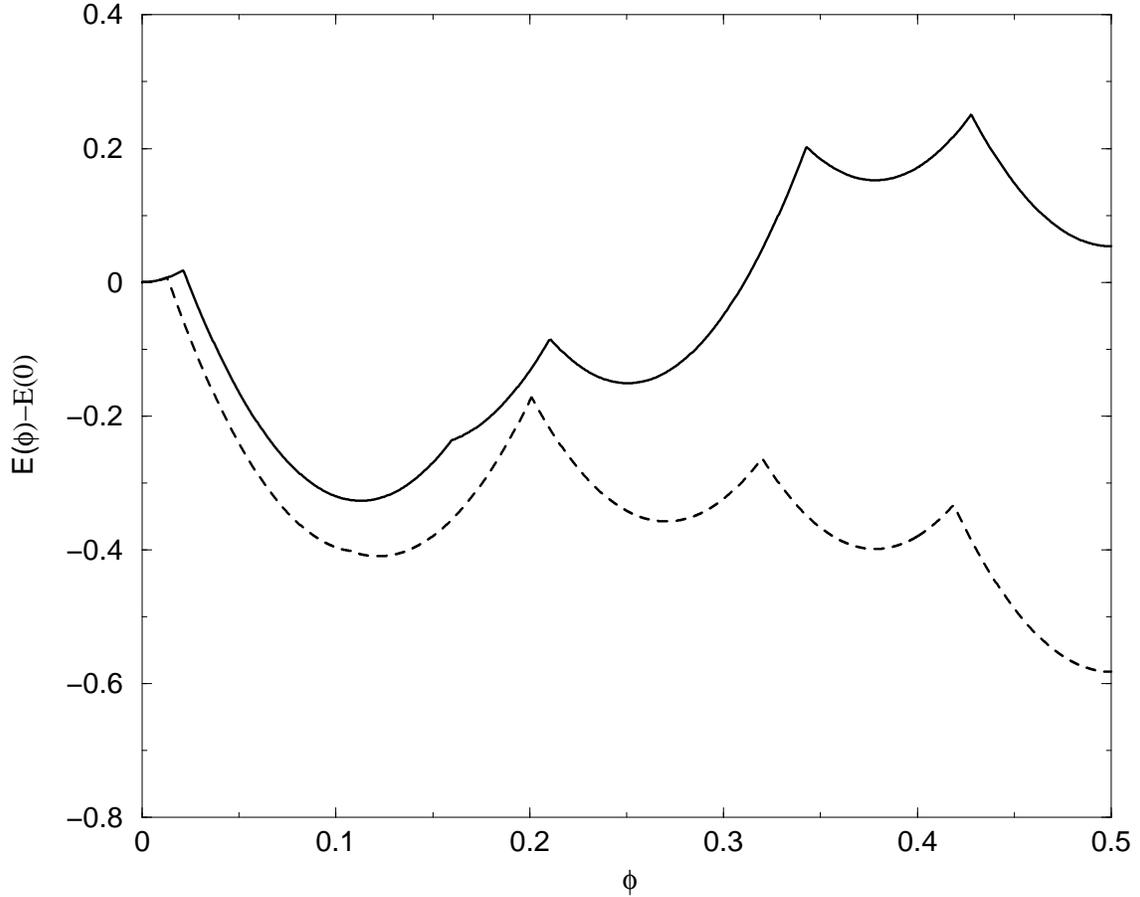,width=15cm,angle=-90}}
\caption{Ground state energy as function of the magnetic flux for a cylinder with $N=M=10$, $N_e=80$ and $S_z=0$. The full curve is obtained with the two reference ansatz (Eq. (\ref{ansatz})) for $U=0.08$ in units of $|t|$, the dashed curve is for the case without interaction ($U=0$). All the cusps are shifted to higher magnetic flux by the long-range Coulomb interaction; this is the Coulomb effect, particularly important in the FS case. One may notice that the same calculation, but with an on-site interaction only (Hubbard model), gives approximately the same result than the non-interacting one.}
\end{figure}

\begin{figure}
\centerline{\psfig{figure=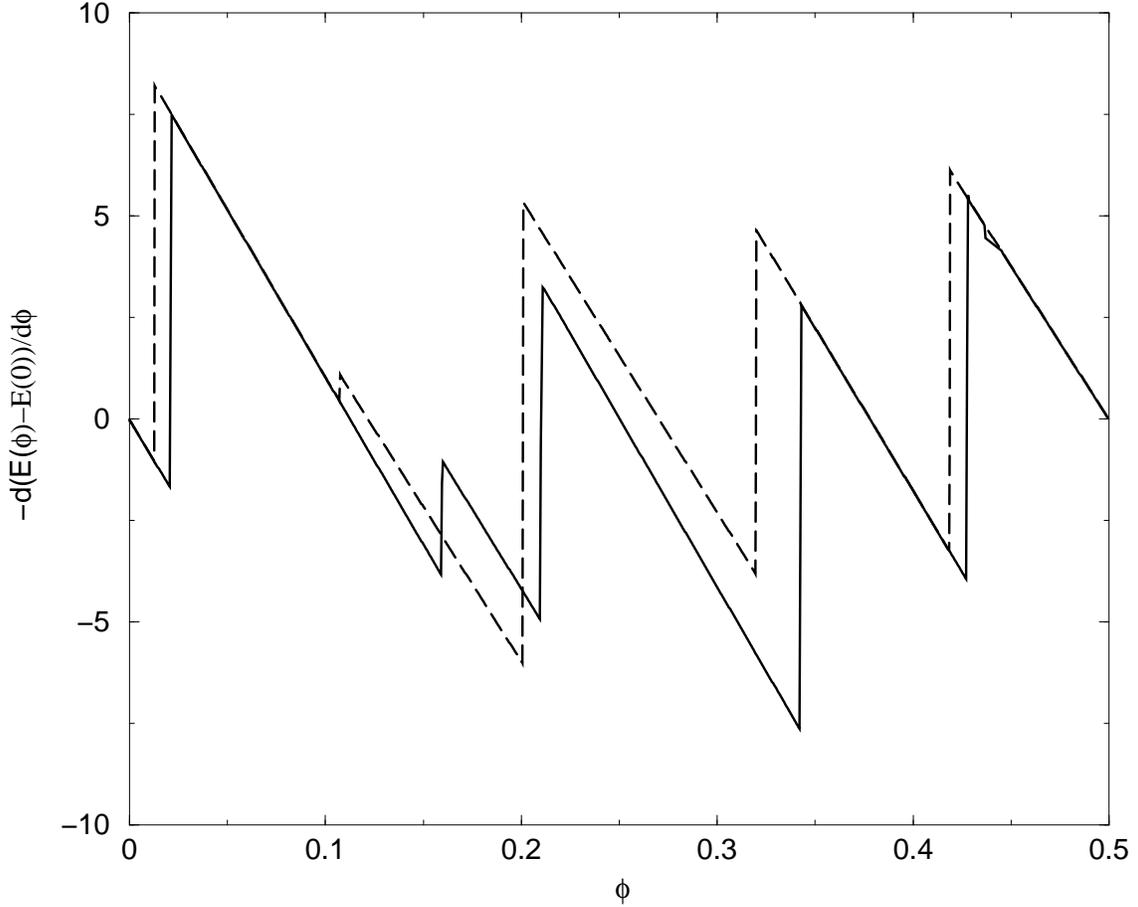,width=15cm,angle=-90}}
\caption{Persistent current as function of the magnetic flux for a cylinder with $N=M=10$, $N_e=80$ and $S_z=0$. The full curve is obtained with the two reference ansatz (Eq. (\ref{ansatz})) for $U=0.08$ in units of $|t|$, the dashed curve is for the case without interaction ($U=0$). The discontinuities are all shifted to higher magnetic flux by the long-range Coulomb interaction. For the FS case, the shift is very pronounced and the importance of the discontinuity is enhanced compared to the non-interacting case. }
\end{figure}

\begin{figure}
\centerline{\psfig{figure=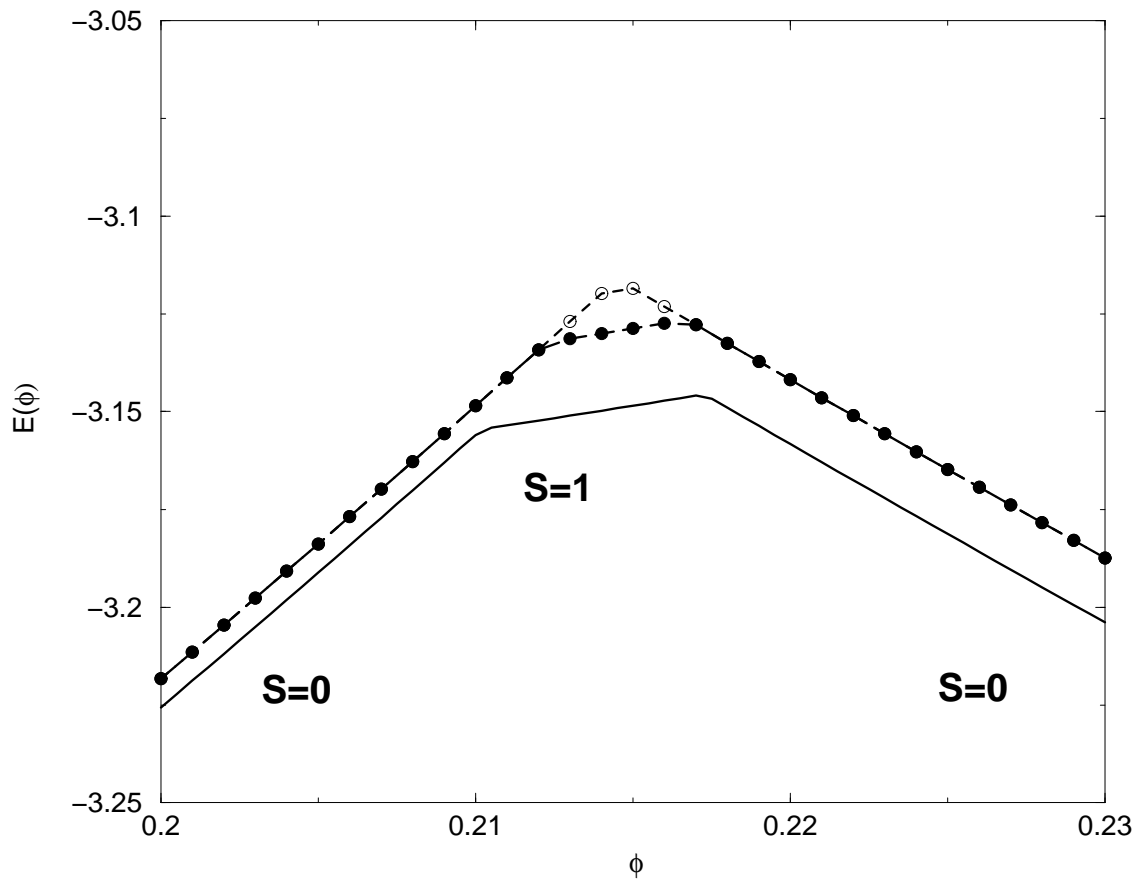,width=15cm,angle=-90}}
\caption{Zoom of the ground state energy of a small cylinder with $N=3$, $M=2$, $N_e=4$ and $S_z=0$, with long range Coulomb interaction, around the crossing point region. The full curve is the exact result for  $U=|t|$. The open circles denote the result obtained with the two reference ansatz (Eq. (\ref{ansatz})). The full dots denote the result obtained by taking the minimum between the values given by the two reference ansatz and the ones given by the triplet built out of the variational solution (Eq. (\ref{energieT})). This approximate results suggest transitions of the total spin, $S=0 \rightarrow S=1 \rightarrow S=0$. The Coulomb interaction, by changing the total spin, induces new real crossing points.}
\end{figure}

\begin{figure}
\centerline{\psfig{figure=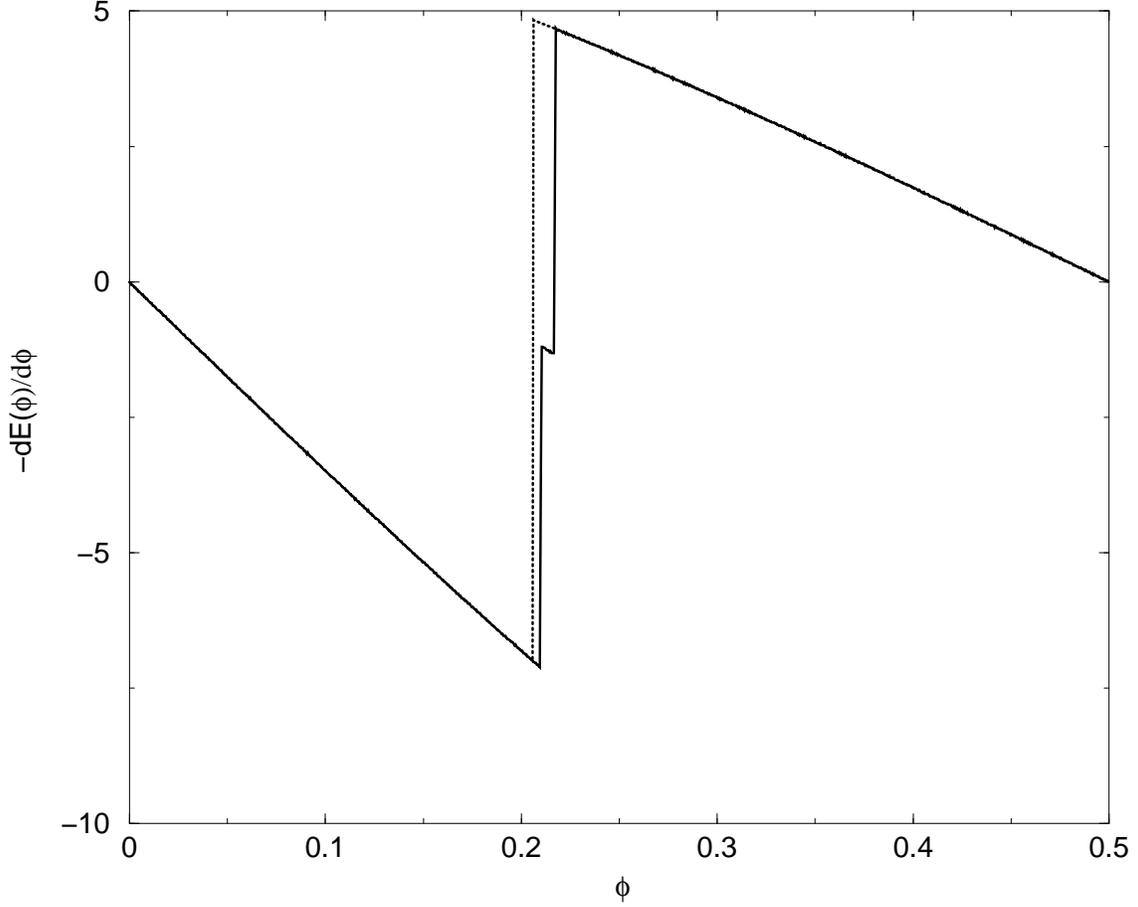,width=15cm,angle=-90}}
\caption{Persistent current as function of the magnetic flux for a cylinder with $N=3$, $M=2$, $N_e=4$ and $S_z=0$. The full curve is the exact result for $U=|t|$ and the dotted curve is for the non-interacting case. The long range Coulomb interaction induces a shift of the discontinuity plus appearance of a new plateau. This plateau can be explained by singlet-triplet transitions (see text).}
\end{figure}

\begin{figure}
\centerline{\psfig{figure=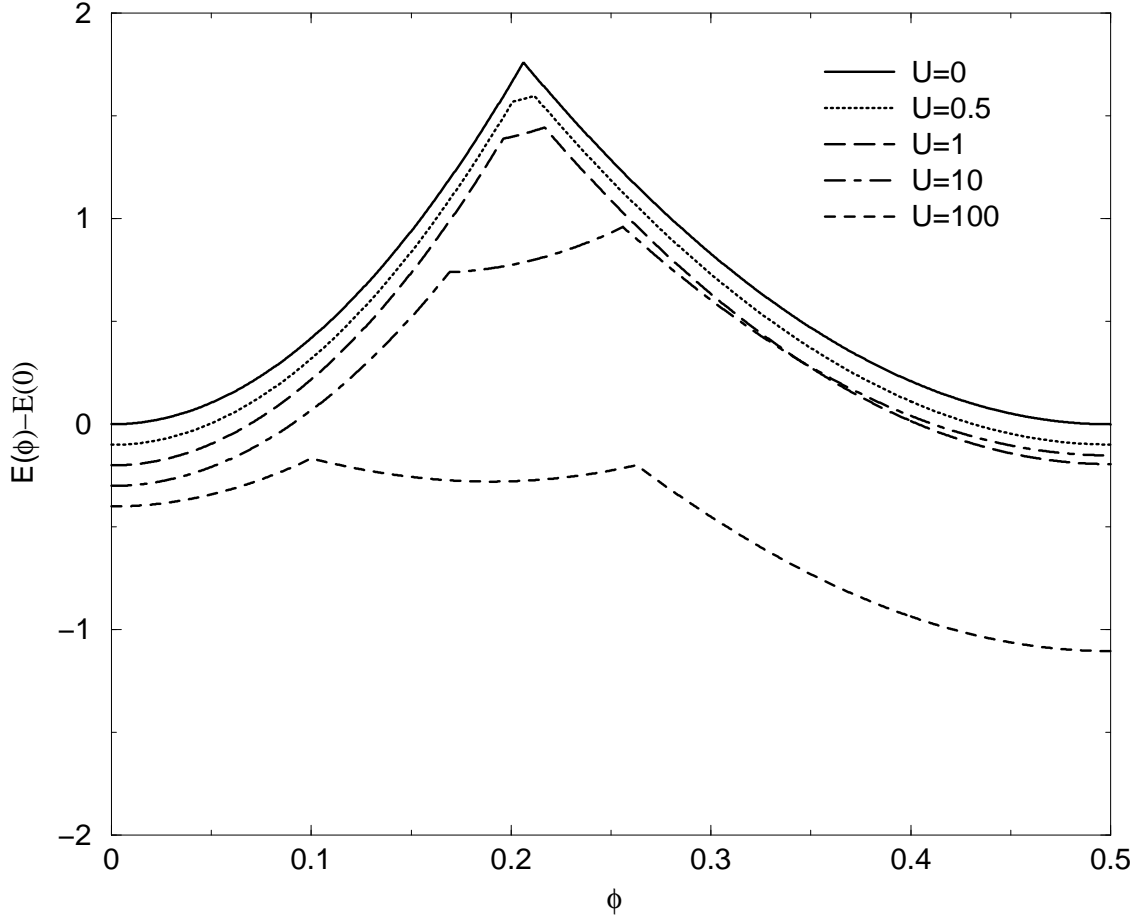,width=15cm,angle=-90}}
\caption{Ground state energy of a small cylinder with $N=3$, $M=2$, $N_e=4$ and $S_z=0$, with short range interaction (Hubbard model) for increasing values of $U$, in units of $|t|$, obtained with exact diagonalisation. Compared to the results with a long-range potential (Fig. 5), the Coulomb effect is strongly reduced but the spin effect is enhanced, in the sense that it becomes relevant for lower values of $U$ and, that the {\it Triplet} plateau is larger. The curves are shifted down for clarity }
\end{figure}

\begin{figure}
\centerline{\psfig{figure=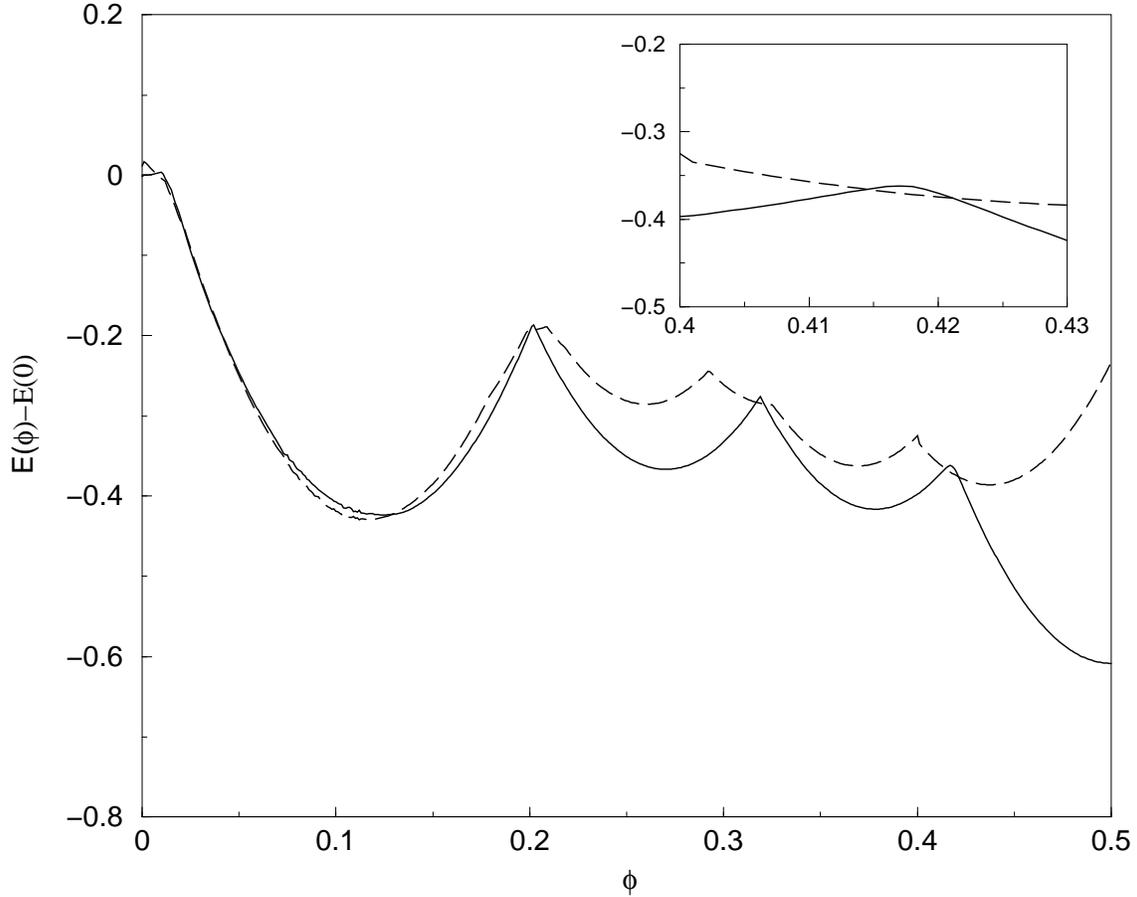,width=15cm,angle=-90}}
\caption{Ground state energy as function of the magnetic flux for a cylinder with $N=M=10$, $N_e=80$ and $S_z=0$ with a short range electron-electron interaction (Hubbard model) and  $U=0.4|t|$. The full curve is the result given by the two reference ansatz (Eq. (\ref{ansatz})), the dashed curve is the energy of the triplet state (Eq. (\ref{energieT})) built out of the variational ansatz. Around each avoided crossing, there is a sequence {\it Singlet} $\rightarrow$ {\it Triplet} $\rightarrow$ {\it Singlet}: for some intervals of magnetic flux, the system is in a triplet ground state. This interval is very large in the vicinity of the FS point. These changes of total spin give new real crossing points. On the one hand, the Coulomb interaction replaces the crossings of the free electron description by avoided crossings, but, on the other hand, creates new crossing points due to spin transition. A zoom on the last crossing points is shown in the inset.}
\end{figure}

\begin{figure}
\centerline{\psfig{figure=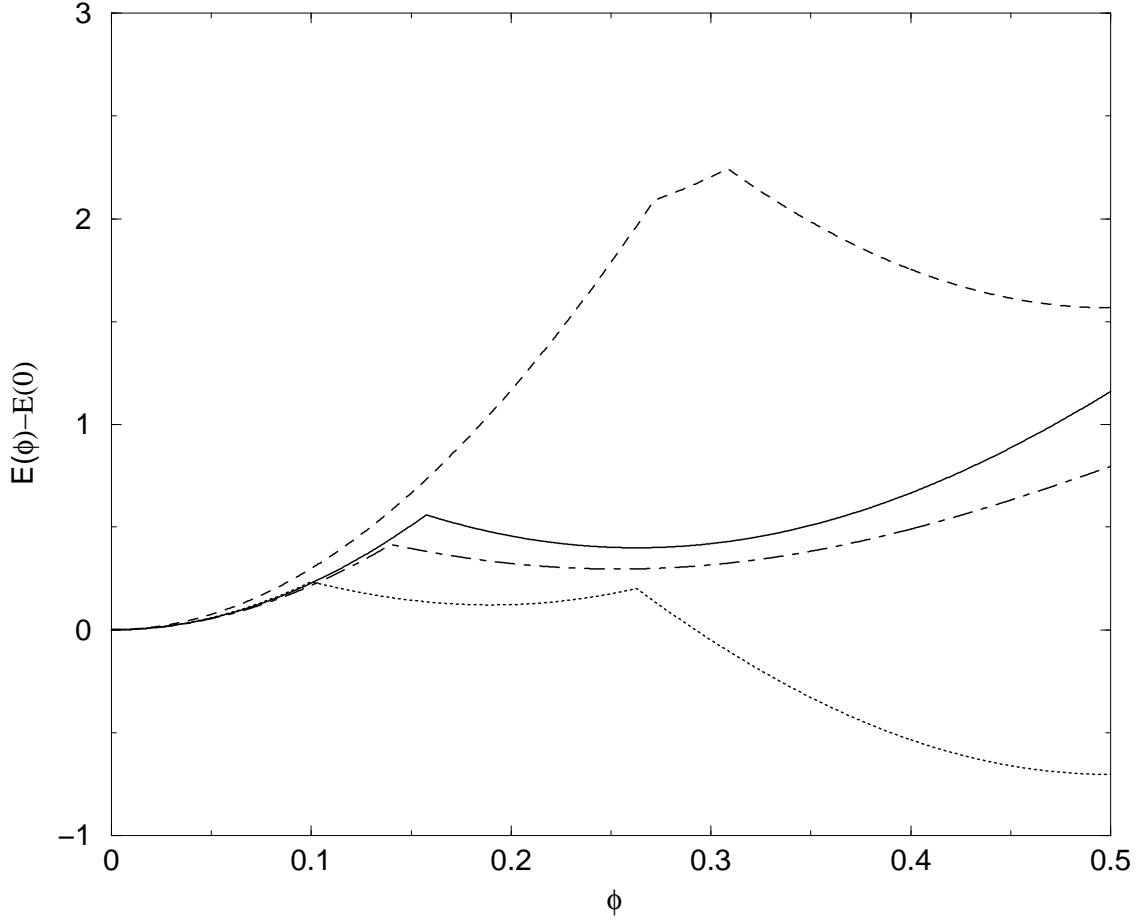,width=15cm,angle=-90}}
\caption{Ground state energy of a small cylinder with $N=3$, $M=2$ and  $U=100|t|$. The dotted and dashed curves are for $N_e=4$, $S_z=0$, short and long range Coulomb potential, respectively. The dot-dashed and full curves are for $N_e=3$, $S_z=1/2$, short and long range Coulomb potential, respectively. The spin effect doesn't occur in the case with one unpaired electron (see text).}
\end{figure}

\end{document}